\documentclass[11pt]{article}
\usepackage{aaspp4,lscape,astrobib}

\let\oldfootsep=\footnotesep
\def\spose#1{\hbox to 0pt{#1\hss}}
\def\simlt{\mathrel{\spose{\lower 3pt\hbox{$\mathchar"218$}}
     \raise 2.0pt\hbox{$\mathchar"13C$}}}
\def\simgt{\mathrel{\spose{\lower 3pt\hbox{$\mathchar"218$}}
     \raise 2.0pt\hbox{$\mathchar"13E$}}}

\def\kmskpc{\,{\rm km \, s^{-1} \, kpc^{-1}}}

\def\msun{{\rm M}_\odot}

\def\Rehat{\widehat{R}_{\rm E}}

\def\umin{u_{\rm min}}

\def\vperp{v_{\rm \perp}}

\def\umin{u_{\rm min}}
\def\t0{t_{\rm 0}}

\def\delchi{\Delta\chi^2}

\def\mlens {M_l} 
 
\def\m1 {M_1} 
\def\m2 {M_2}

\def\Ebv{E(B-V)}

\def\vhat{\widehat{v}} 
\def\that {\widehat{t}}
 
\def\kms {\,{\rm km \, s^{-1} }}
\def\kpc {\, {\rm kpc}} 

\begin{document}
\title{MACHO 96-LMC-2: Lensing of a Binary Source in the LMC and
  Constraints on the Lensing Object}
\author{
      C.~Alcock\altaffilmark{1,2},
    R.A.~Allsman\altaffilmark{3},
      D.R.~Alves\altaffilmark{4},
    T.S.~Axelrod\altaffilmark{5},
      A.C.~Becker\altaffilmark{6,7},
    D.P.~Bennett\altaffilmark{8},
    K.H.~Cook\altaffilmark{1,2},
    A.J.~Drake\altaffilmark{1,5},
    K.C.~Freeman\altaffilmark{5},
      M.~Geha\altaffilmark{1,9},
      K.~Griest\altaffilmark{2,10},
    M.J.~Lehner\altaffilmark{11},
    S.L.~Marshall\altaffilmark{1,2},
    D.~Minniti\altaffilmark{1,12},
    C.A.~Nelson\altaffilmark{1,13},
    B.A.~Peterson\altaffilmark{5},
      P.~Popowski\altaffilmark{1},
    M.R.~Pratt\altaffilmark{14},
    P.J.~Quinn\altaffilmark{15},
    C.W.~Stubbs\altaffilmark{2,6},
      W.~Sutherland\altaffilmark{16},
    A.B.~Tomaney\altaffilmark{6},
      T.~Vandehei\altaffilmark{2,10},
      D.~Welch\altaffilmark{17}
        }
\begin{center}
{\bf (The MACHO Collaboration) }\\
\end{center}

%-------------------- institutions and email ----------------------

\altaffiltext{1}{Lawrence Livermore National Laboratory, Livermore, CA 94550\\
  Email: {\tt alcock, dminniti, kcook, adrake, mgeha, cnelson, popowski, stuart@igpp.ucllnl.org}}

\altaffiltext{2}{Center for Particle Astrophysics, University of California, Berkeley, CA 94720}

\altaffiltext{3}{Supercomputing Facility, Australian National University,
  Canberra, ACT 0200, Australia \\
  Email: {\tt Robyn.Allsman@anu.edu.au}}

\altaffiltext{4}{Space Telescope Science Institute, 3700 San Martin Dr.,
  Baltimore, MD 21218\\
  Email: {\tt alves@stsci.edu}}

\altaffiltext{5}{Research School of Astronomy and Astrophysics, Australian National University, 
  Canberra, ACT 2611, Australia\\
  Email: {\tt tsa, kcf, peterson@mso.anu.edu.au}}

\altaffiltext{6}{Departments of Astronomy and Physics,
  University of Washington, Seattle, WA 98195\\
  Email: {\tt austin, stubbs@astro.washington.edu}}

\altaffiltext{7}{Bell Laboratories, Lucent Technologies, 600 Mountain Avenue, Murray Hill, NJ 07974\\
  Email: {\tt acbecker@physics.bell-labs.com}}

\altaffiltext{8}{Department of Physics, University of Notre Dame, IN 46556\\
  Email: {\tt bennett@bustard.phys.nd.edu}}

\altaffiltext{9}{Department of Astronomy and Astrophysics, University of California, Santa Cruz,  CA 95064}

\altaffiltext{10}{Department of Physics, University of California,
  San Diego, CA 92039\\
  Email: {\tt kgriest@ucsd.edu, vandehei@astrophys.ucsd.edu }}

\altaffiltext{11}{Department of Physics, University of Sheffield, Sheffield S3 7RH, UK\\
  Email: {\tt m.lehner@sheffield.ac.uk}}

\altaffiltext{12}{Depto. de Astronomia, P. Universidad Catolica, Casilla 104,
  Santiago 22, Chile\\
  Email: {\tt dante@astro.puc.cl}}

\altaffiltext{13}{Department of Physics, University of California, Berkeley,  CA 94720}

\altaffiltext{14}{Center for Space Research, MIT, Cambridge, MA 02139\\
  Email: {\tt mrp@ligo.mit.edu}}

\altaffiltext{15}{European Southern Observatory, Karl Schwarzchild Str.\ 2, D-8574 8 G\"{a}rching bei M\"{u}nchen, Germany\\
  Email: {\tt pjq@eso.org}}

\altaffiltext{16}{Department of Physics, University of Oxford,
  Oxford OX1 3RH, U.K.\\
  Email: {\tt w.sutherland@physics.ox.ac.uk}}

\altaffiltext{17}{McMaster University, Hamilton, Ontario Canada L8S 4M1\\
  Email: {\tt welch@physics.mcmaster.ca}}

\setlength{\footnotesep}{\oldfootsep}
\renewcommand{\topfraction}{1.0}
\renewcommand{\bottomfraction}{1.0}     % allows big pictures on the bottom

\newpage

\begin{abstract} 
\rightskip = 0.0in plus 1em

We present photometry and analysis of the microlensing alert MACHO
96-LMC-2 (event LMC-14 in \citeNP{macho-lmc5}).  This event was
initially detected by the MACHO Alert System, and subsequently
monitored by the Global Microlensing Alert Network (GMAN).  The $\sim
3\%$ photometry provided by the GMAN follow--up effort reveals a
periodic modulation in the lightcurve.  We attribute this to binarity
of the lensed source.  Microlensing fits to a rotating binary source
magnified by a single lens converge on two minima, separated by
$\delchi \sim 1$.  The most significant fit X1 predicts a primary
which contributes $\sim 100 \%$ of the light, a dark secondary, and an
orbital period ($T$) of $\sim 9.2$ days.  The second fit X2 yields a
binary source with two stars of roughly equal mass and luminosity, and
$T = 21.2$ days.

Observations made with the Hubble Space Telescope
(HST)\footnote[1]{The NASA/ESA Hubble Space Telescope is operated by
AURA, Inc., under NASA contract NAS5--26555.}  resolve stellar
neighbors which contribute to the MACHO object's baseline brightness.
The actual lensed object appears to lie on the upper LMC main
sequence.
%The MACHO 96-LMC-2 source's blue color and location on the main
%sequence imply it is unlikely that it is located far behind the LMC;
%thus it serves as an important counter-example to suggestions that LMC
%self-lensing arises primarily from a very extended distribution of
%source stars along the line of sight (e.g., \citeNP{zhao-extinct}).
%We use \citeN{girardi2000} isochrones to estimate the mass of the
%primary component of the binary system, $M \sim 2~\msun$.  This helps
We estimate the mass of the primary component of the binary system, $M
\sim 2~\msun$.  This helps to determine the physical size of the
orbiting system, and allows a measurement of the lens proper motion.
For the preferred model X1, we explore the range of dark companions by
assuming $0.1 \msun$ and $1.4 \msun$ objects in models X1a and X1b,
respectively.  We find lens velocities projected to the LMC in these
models of $\vhat_{X1a} = 18.3 \pm 3.1 \kms$ and $\vhat_{X1b} = 188 \pm
32 \kms$.  In both these cases, a likelihood analysis suggests an LMC
lens is preferred over a Galactic halo lens, although only marginally
so in model X1b.  We also find $\vhat_{X2} = 39.6 \pm 6.1 \kms$, where
the likelihood for the lens location is strongly dominated by the LMC
disk.  In all cases, the lens mass is consistent with that of an
M-dwarf.  Additional spectra of the lensed source system are necessary
to further constrain and/or refine the derived properties of the
lensing object.

The LMC self-lensing rate contributed by 96-LMC-2 is consistent with
model self-lensing rates.  Thus, even if the lens is in the LMC disk,
it does not rule out the possibility of Galactic halo microlenses
altogether.  Finally, we emphasize the unique capability of follow-up
spectroscopic observations of known microlensed LMC stars, combined
with the non-detection of binary source effects, to locate lenses in
the Galactic halo.

%The LMC self-lensing rate contributed by events MACHO 96-LMC-2 and
%possible LMC lens MACHO LMC-9 \cite{macho-binary} is consistent with
%the model LMC self-lensing rates calculated in \citeN{macho-lmc5}.  We
%caution this event does not rule out halo microlenses altogether, as
%might be construed from \citeN{kerins-evans}.  Finally, we emphasize
%the unique capability of follow-up spectroscopic observations of known
%microlensed LMC stars, combined with the non-detection of binary source
%effects during peak magnification, to locate lenses in the Galactic
%halo.

\end{abstract}
\vspace{-5mm}
\keywords{dark matter - gravitational lensing - stars: low-mass, brown
dwarfs - binaries: general}

\newpage
\section{Introduction}
\label{sec-intro}
 
The interpretation of gravitational microlensing results towards the
Magellanic Clouds (e.g., \citeNP{eros-smc2,macho-lmc5}) has been
hindered by the unknown location of the lensing systems.  Exceptions
to this are caustic crossing binary lens events MACHO LMC-9
\cite{macho-binary}, where a sparsely resolved caustic crossing
suggests an LMC lens, and MACHO 98-SMC-1
(\citeNP{macho-98smc1,98smc1-all}), where a lens association with the
SMC is more certain.  In this paper we present an additional
``exotic'' microlensing event seen towards the Magellanic Clouds,
MACHO 96-LMC-2.
 
The lightcurve of MACHO 96-LMC-2 exhibits deviations from the standard
microlensing fit similar to those that are expected if a binary source
is lensed \cite{griest-hu92}.  This effect is the ``inverse'' of the
parallax effect due to motion of the Earth around the Sun
\cite{gould92par,macho-par}, and may be referred to as the
``xallarap'' effect, where the orbital motion occurs at the lensed
source.  Detection of the xallarap modulation in a microlensing
lightcurve allows us to fit the semi-major axis of the orbiting system
in units of the lens' projected Einstein ring radius.  An estimate of
the physical semi-major axis of the system then allows a second
constraint (along with the event timescale $\that$) on the 3
degenerate lens parameters mass, distance, and transverse velocity.
\citeN{han97a} describe the use of this effect in discriminating
between Galactic halo and LMC lenses.

A detection of this type of modulation in a microlensing lightcurve is
possible within a certain range of event parameters.  First, the
binary source should have an orbital period similar to or shorter than
the event timescale, such that the sources accelerate appreciably
during the time they are microlensed and an orbital period may be
determined.  Second, the orbital separation of the sources should not
be much smaller than the lens' Einstein ring radius projected to the
source, otherwise the system appears to the lens as essentially a
single object.  This biases the detection of the xallarap effect
towards events where the lens is close to the sources, analogous to
the parallax effect, which is most easily detected when the lens is
relatively close to the Sun-Earth system.  A comprehensive review of
microlensing with rotating binaries (lenses, sources, and observers)
may be found in \citeN{dominik98a}.

\section{Observations}
\label{sec-observations}
 
\begin{figure}
\plotone{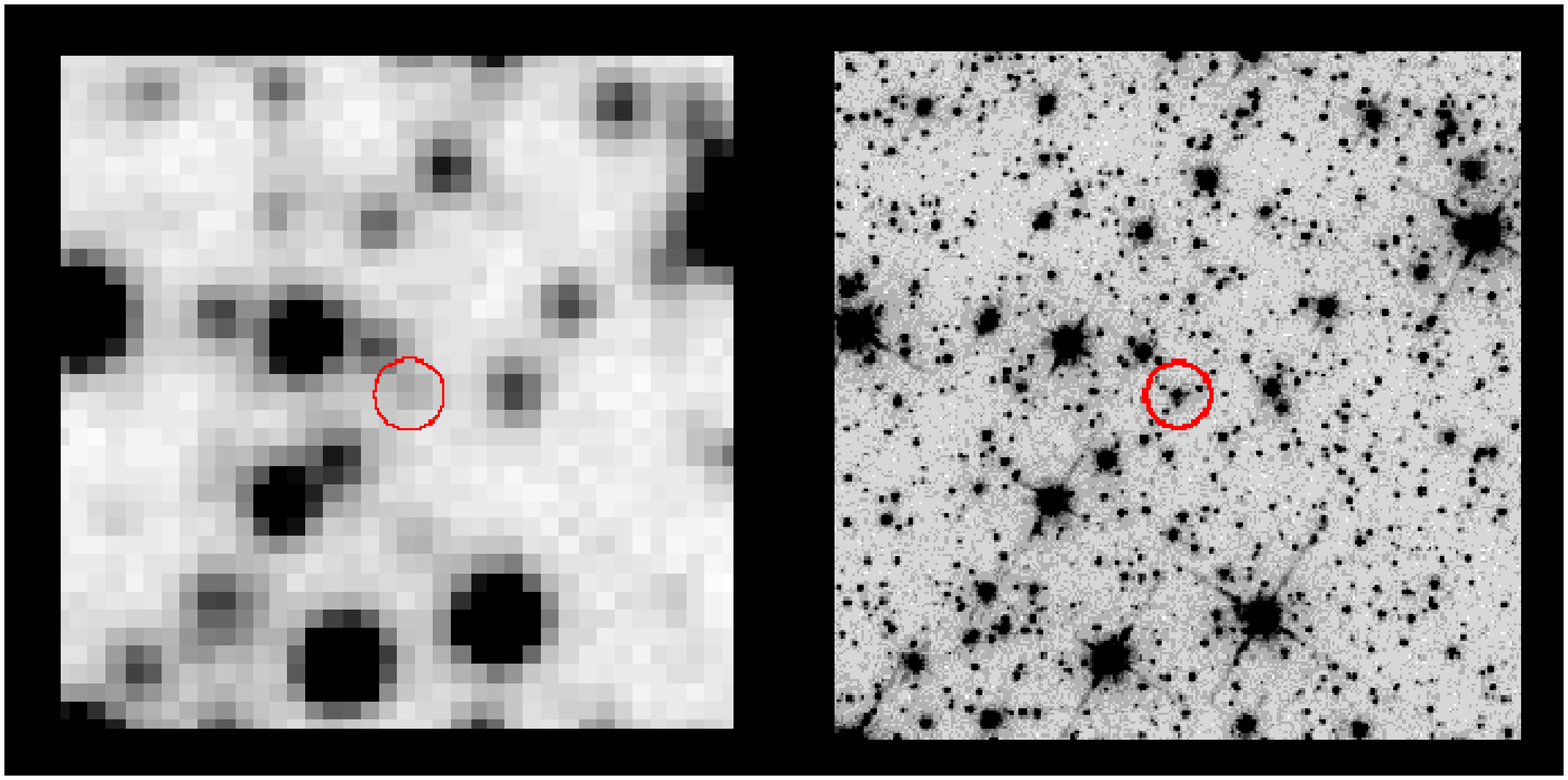}
\figcaption[f1.ps]{\label{fig-image}
  Each image represents a $25\arcsec \times 25\arcsec$ field centered
on the lensed source object.  The image on the left is from the 300
sec MACHO R-band template observation, and on the right is a 4 x 500
sec combined HST R-band image.  The lensed source is in the center of
each image, and is indicated by the circle.  North is up and east is
to the left.  }
\end{figure}

Microlensing Alert MACHO 96-LMC-2 was detected and announced on Oct 3,
1996, with the MACHO object at an observed magnification of $A \sim
1.8$.  The source for this event is located at $\alpha = $
05:34:44.437, $\delta = -$70:25:07.37 (J2000), in the south-east
extreme of the LMC bar.  This object was constant at $V = 19.42 \pm
0.15, (V-R) = -0.03 \pm 0.10$ in $\sim 700$ observations over the 4.2
years preceding this brightening, where these magnitude errors are
dominated by uncertainty in the calibration of the MACHO database
\cite{macho-calib}.  The MACHO ID number for this star is
$11.8871.2108$.  A $25\arcsec \times 25\arcsec$ postage stamp, taken
from the MACHO R-band template observation of this field and centered
around the lensed object, is presented in Figure~\ref{fig-image}.

Nightly observations were requested on the CTIO\footnote[2]{Cerro
Tololo Interamerican Observatory, National Optical Astronomy
Observatories, operated by the Association of Universities for
Research in Astronomy, Inc. under cooperative agreement with the
National Science Foundation.} 0.9m telescope as part of the Global
Microlensing Alert Network (GMAN) microlensing follow-up program
(\citeNP{macho-apj97e,macho-98smc1,macho-binary}).  Data were obtained
in both $B$ and Kron-Cousins $R$ for the duration of this event.
Final sets of baseline observations were made $\sim 800$ days ($\sim
8$ times the duration of the event) after the peak.  An additional set
of observations was made at the UTSO 0.6m telescope.  However, closure
of the telescope in 1997 prevented baseline measurements from this
site.

After the event, cycle-7 HST images were taken with the target star
centered in the Planetary Camera of HST instrument WFPC2.  These
integrations included four 500 second exposures in each of three bands
$V$, $R$, and $I$.  The images were combined using the IRAF routine
imcombine, along with a sigma clipping algorithm to remove cosmic
rays.  Aperture photometry was performed on all stars using the
DaoPhot package \cite{stetson94}, with centroids derived from a PSF
fit.  We used a 0.25$\arcsec$ aperture, and corrected to a
0.5$\arcsec$ aperture using the brightest stars in the field.  We
corrected for the charge transfer effect and calibrate the magnitudes
using the \citeN{wfpc-calib} calibrations.  A portion of the combined
R-band image of the WFPC2-imaged field around event 96-LMC-2 is
presented in Figure~\ref{fig-image}.

\begin{figure}
\plotone{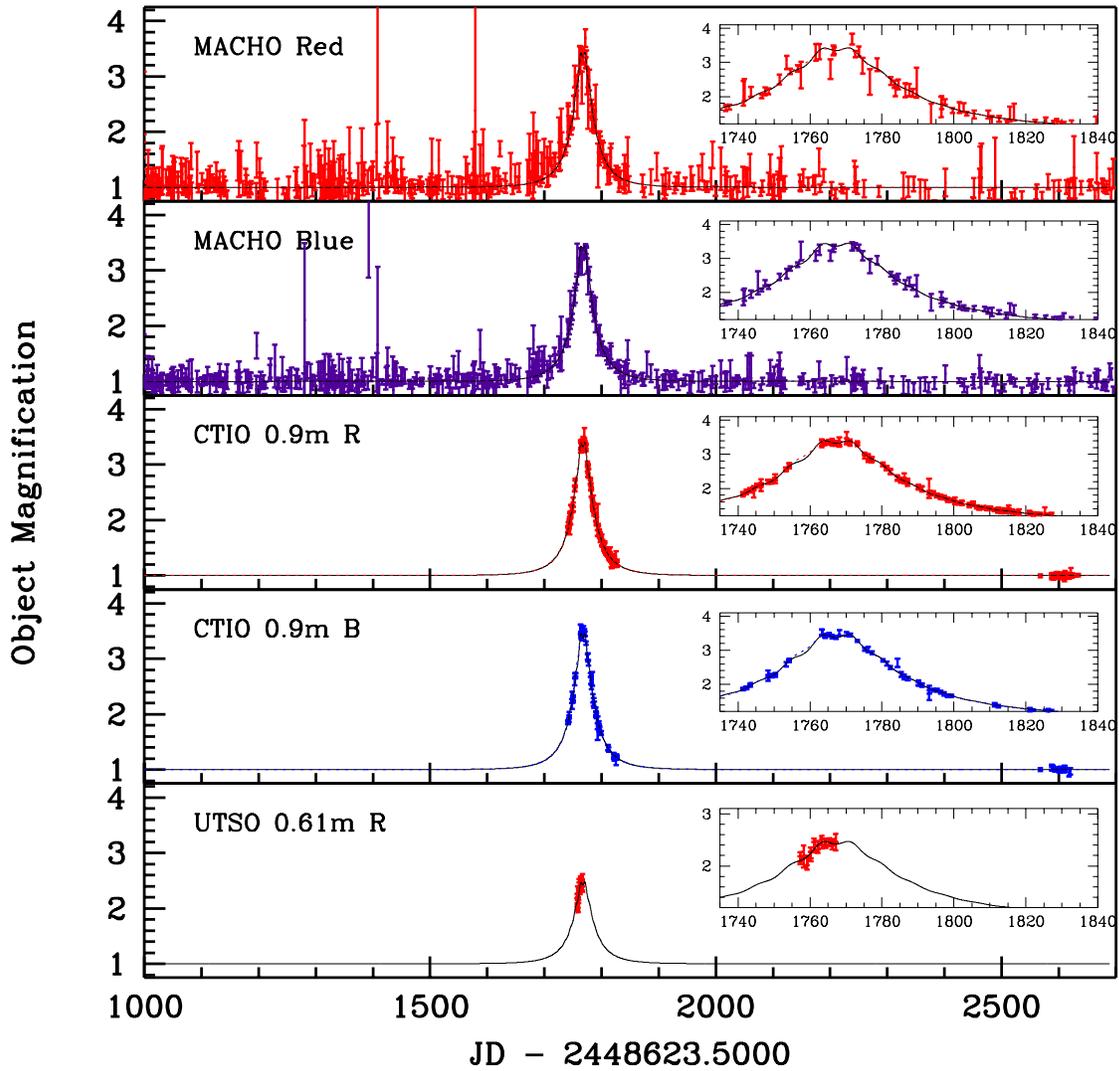}
\figcaption[f2.ps]{\label{fig-fit}
  The light curve of event 96-LMC-2. The panels show the observed
brightening as a function of time, with passbands and sites as
indicated.  The solid line represents the best ``xallarap'' fit X1,
and the dotted line fit X2.  Insets are provided to better view the
region with the strongest binary source signal.  The UTSO data are
only plotted with the X1 fit curve. }
\end{figure}

The MACHO/GMAN data for 96-LMC-2 are presented in
Figure~\ref{fig-fit}.  The MACHO data were reduced with MACHO's
standard photometry package SoDOPHOT, with minimum errors of 0.014
added in quadrature.  The CTIO and UTSO data were simultaneously
reduced with the ALLFRAME package \cite{stetson-allframe}, and the
error estimates are multiplied by a factor of 1.5 to account for
global systematics (such as flat-fielding errors and the swapping of
CCD detectors) in the time series of data.

\section{Microlensing Fits}
\label{sec-fits}

\subsection{Standard Microlensing}

Standard microlensing fit parameters, including the effects of
unlensed contributions to the source baseline flux, are presented in
Table~\ref{tab-stdparams}.  These include

\begin{itemize}

\item $\t0$, the time of closest approach of lens to source-observer
  line of sight,
\item $\that$, the Einstein ring diameter crossing time,
\item $\umin$, the lens impact parameter in units of the Einstein ring
  radius.

\end{itemize}
 
\begin{figure}
\plotone{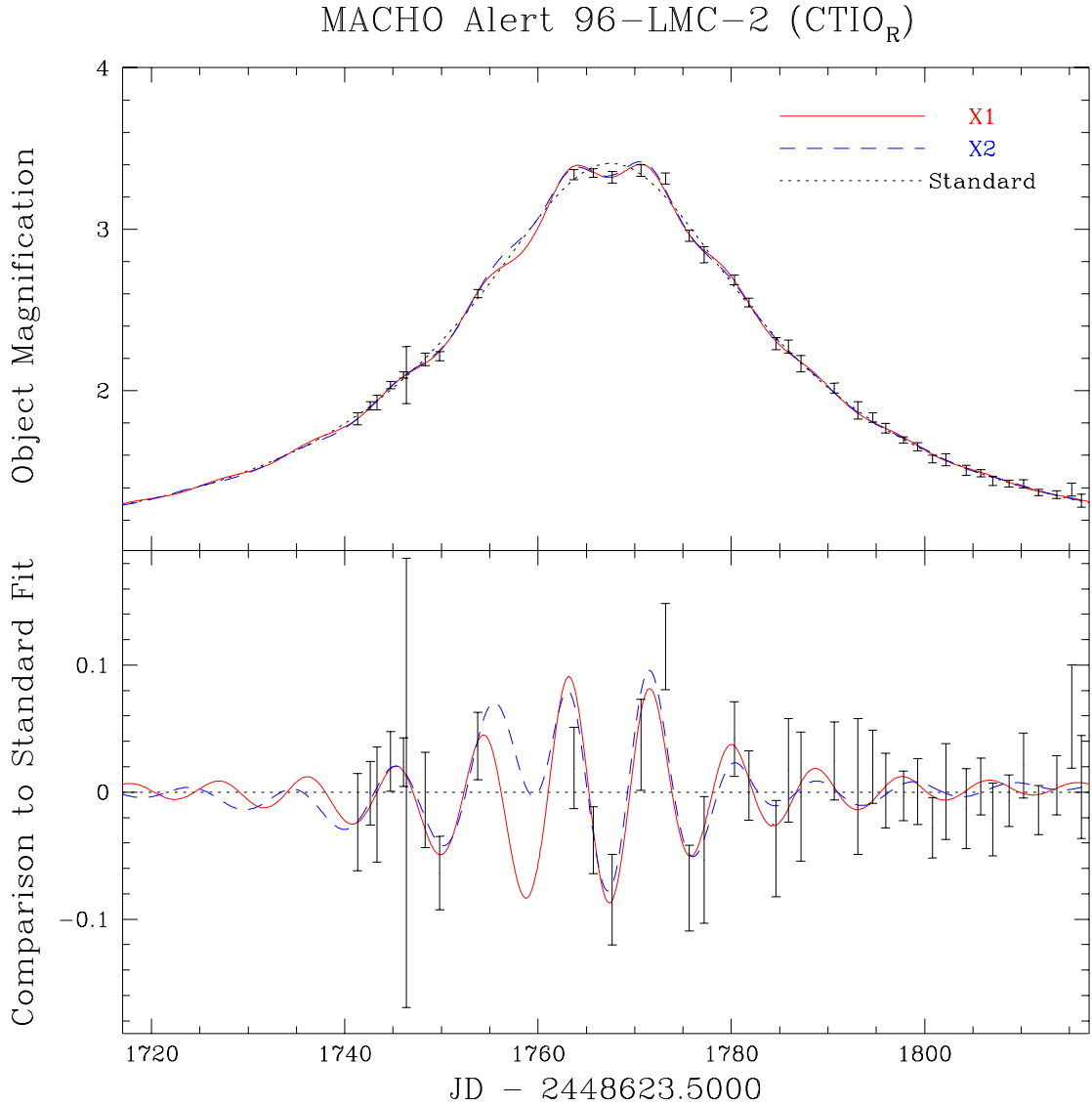}
\figcaption[f3.ps]{\label{fig-residr}
  The CTIO R-passband light curve of 96-LMC-2. The top panel displays
the 3 fits performed on the complete time-series of data, along with
the CTIO R-band data in 1-day bins.  The dotted line represents the
best standard fit, and the solid and dashed lines the most significant
and second most significant xallarap fits X1 and X2, respectively.
The bottom panel indicates the residuals of each fit and the data
around the standard microlensing fit.}
\end{figure}

\begin{figure}
\plotone{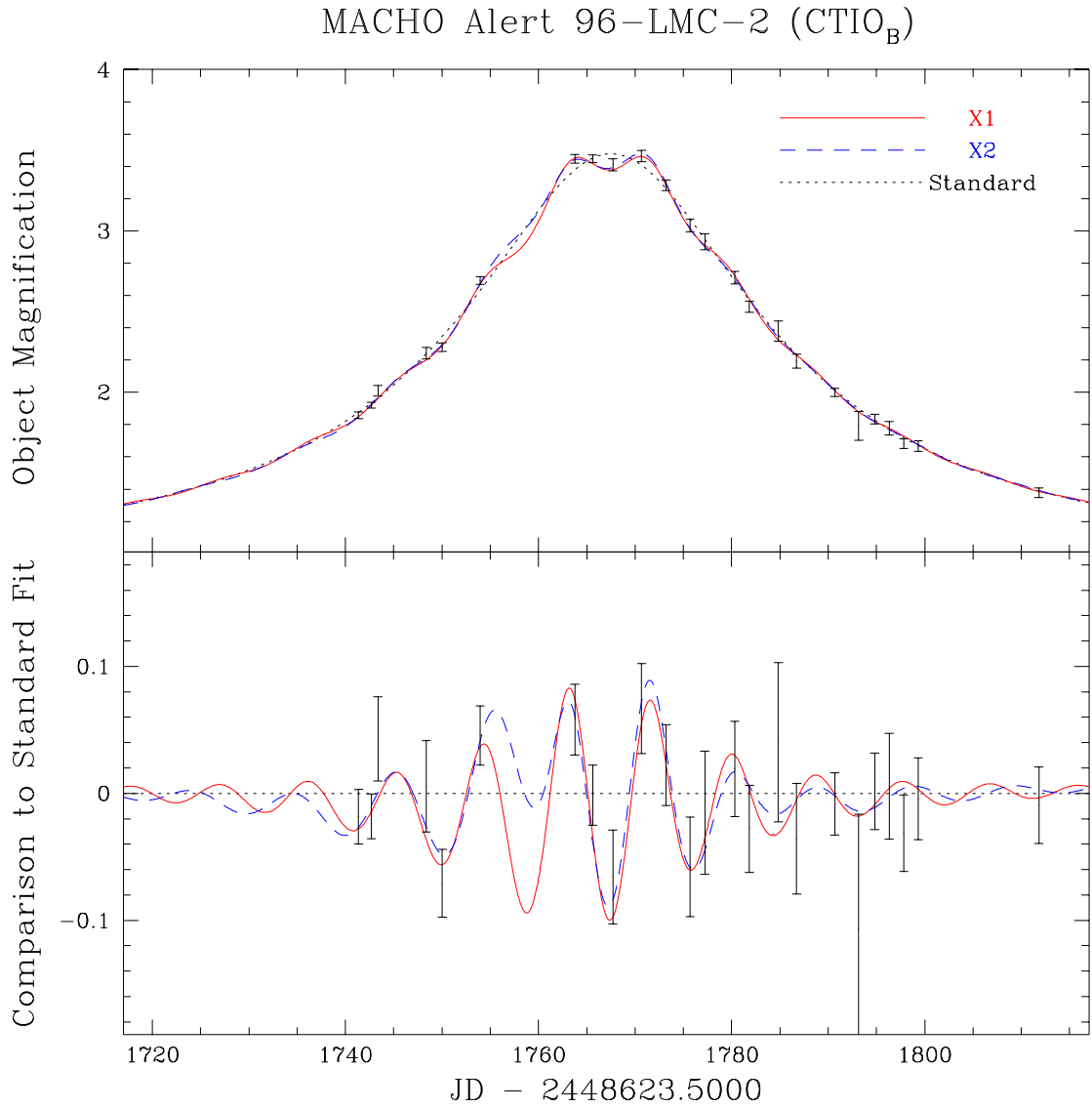}
\figcaption[f4.ps]{\label{fig-residb}
  Same as Figure~\ref{fig-residr}, except with the CTIO B-band light
curve.}
\end{figure}

We also include 5 additional blending parameters, one $f$ for each
passband of observations.  In each case, $f$ represents the fraction
of the object's baseline flux which was lensed.  The $\chi^2 / d.o.f$
for this fit is 0.91, formally an acceptable fit.  However, there are
periodic residuals around this smooth fit, especially in the CTIO
data.  We have plotted these residuals for the CTIO R and B passbands
in Figure~\ref{fig-residr} and Figure~\ref{fig-residb}, respectively.

\subsection{Binary Source (Xallarap) Microlensing}
\label{subsec-xal}

Fitting this event to an orbiting binary source does provide
significant improvement over the standard blended fit.  We follow the
formalism of \citeN{dominik98a} for the binary source solution.  In
particular, we use the fit parameters:

\begin{itemize}

\item $\tilde{t}_b$, the time of closest approach of the lens to the source system center of mass,
\item $t_{\rm E}$, the lens' Einstein radius crossing time,
\item $\tilde{b}$, the lens' impact parameter with respect to the source
  system center of mass, in units of the lens' Einstein radius projected into
  the source plane,
\item $\tilde{\alpha}$, the angle between the lens trajectory and the $x$ source axis,
\item $\tilde{f}_1$, the total binary flux fraction of source 1,
\item $\tilde{m}_1$, the total binary mass fraction of source 1,
\item $\tilde{\rho}$, the orbital semi-major axis in units of the lens' projected 
  Einstein radius,
\item $\tilde{\beta}$, the orbital inclination,
%\item $\tilde{\gamma}$, the inclination around the $y$ source axis,
\item $\tilde{T}$, the orbital period in days,
\item $\tilde{\xi}_0$, the orbital phase at time $\tilde{t}_b$,

\end{itemize}

and one $f$ for each passband of observations, an additional 8
parameters compared to the standard blended microlensing fit.
Finally, we assume zero eccentricity circular orbits for the sources,
meaning inclination angle $\tilde{\gamma}$ from \citeN{dominik98a} is
redundant, and is set to zero in these fits.

We find 2 minima in this parameter space, separated by $\delchi \sim
1$.  Our most significant model is labeled X1, and our second most
significant is X2.  These fits are a further $\delchi = 72$ from the
standard microlensing fit, which we are extremely unlikely to arrive
at by chance, even given our additional 7 constraints.  Xallarap fit
parameters are presented in Table~\ref{tab-xalparams}.  Fit X1
indicates a primary which contributes $\sim 100 \%$ of the light, a
dark secondary, and an orbital period of $T = 9.22 \pm 0.21$ days.
The second fit X2 yields a binary source of similar mass and
brightness stars, and $T = 21.2 \pm 0.54$ days.  These fits and the
residuals of these fits around the standard microlensing fit are
plotted along with the data for the CTIO R and B passbands in
Figure~\ref{fig-residr} and Figure~\ref{fig-residb}, respectively.

We have investigated whether or not the sources contribute
significantly different flux fractions in the red and blue, by
allowing different $\tilde{f}_1$ fractions for the red and blue
passbands.  Providing this additional constraint leads to a $\delchi =
-0.24 (-0.50)$ for fit X1 (X2), indicating this improvement is
formally significant at only the $37\% (52\%)$ confidence level.  For
the fit X1 class, the secondary source is dark in both red and blue,
to within the reported accuracy of our photometry.  For the X2 class,
the best fit $\tilde{f}_{1r}/\tilde{f}_{1b}$ is 1.02.  In the
following we assume $\tilde{f}_{1r} = \tilde{f}_{1b} = \tilde{f}_1$.

The binary source fits for this event are plotted with the data in
Figure~\ref{fig-fit}.  Due to the different blend fractions for these
fits, the observed object magnification (as opposed to the lensed
source magnification) is plotted as a function of time.  It is
apparent that without the GMAN follow-up photometry, there would be
little support for the binary source interpretation.  In fact, $80 \%$
of the $\delchi$ between the standard fit and fit X1 is contributed by
the GMAN data.

\subsubsection{Colors of the Lensed Objects}

The source object's brightness is well constrained with our $V, R,$
and $I$ HST images.  We use the image subtraction method of
\citeN{tomaney96} to locate the lensed source in the MACHO images to
within 0.1'' (2 PC pixels).  Image registration allows us to uniquely
determine the centroid of the lensed source in the HST image.  This
lensed flux aligns with an object with $V = 19.46 \pm 0.02, (V - R) =
0.00 \pm 0.03, (V - I) = 0.12 \pm 0.03$, where the errors represent
the quadratic sum of Poisson noise and an adopted error of $2\%$ for
our HST magnitudes.  Within $1''$ of this source there are at least 2
neighbors, which contribute about $20\%$ of the flux within this
region.  
%NTT obs indicate $K = 19.72 \pm 0.10$.

\begin{figure}
\plotone{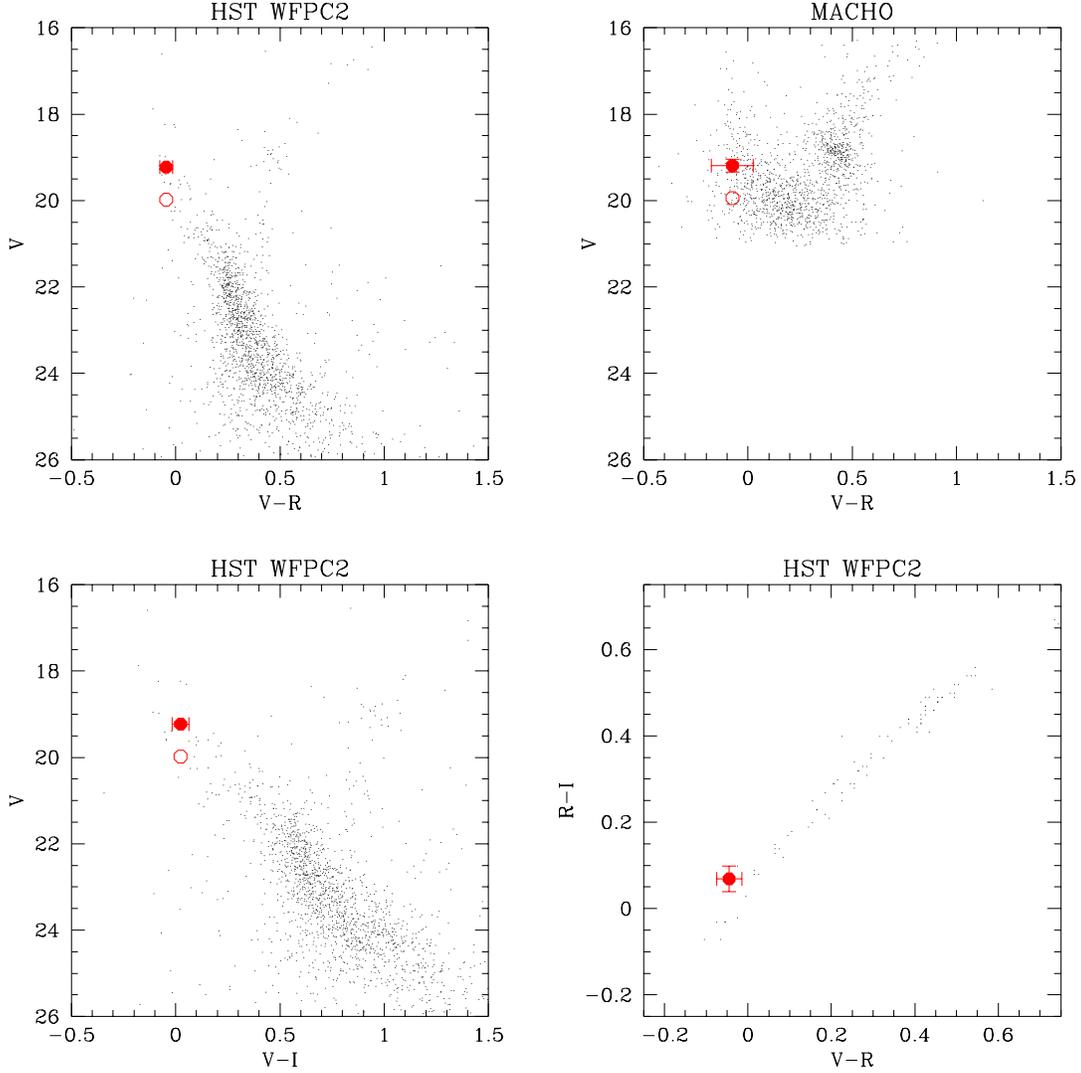}
\figcaption[f5.ps]{\label{fig-cmd}
  Color-magnitude and color-color diagrams of the field surrounding
event 96-LMC-2.  The lensed object is indicated with a filled circle.
The HST CMDs are constructed from cycle-7 observations with the target
star centered on the Planetary Camera chip of the WFPC2.  The MACHO
CMD is taken from a subsection of the template observation of this
field.  Error bars indicate the uncertainties in each respective CMD.
This object resides near the upper main sequence of the LMC field.  We
indicate with an open circle the region of the CMD where 2 equal
brightness stars must be drawn from to create the observed object.
}
\end{figure}

Color-magnitude diagrams (CMDs) incorporating $\sim$1800 objects from
the Planetary Camera chip of the WFPC2, and a region surrounding the
lensed object in the MACHO focal plane, are displayed in
Figure~\ref{fig-cmd}.  The lensed object identified in each respective
CMD is indicated with the filled circle.  We have corrected the CMDs
for reddening, using a characteristic LMC reddening in the bar of
$\Ebv = 0.07 \pm 0.01$ (e.g., \citeNP{knut-99aj} and
\citeNP{holtzman-99}) and the Landolt extinction coefficients of
\citeN{schlegel-98}.  The intrinsic magnitude and colors of the source
star are $V = 19.23 \pm 0.04, (V - R) = -0.04 \pm 0.03, (V
- I) = 0.02 \pm 0.03$. %, (I - K) = -0.43 \pm 0.10.

%$V = 19.31 \pm 0.04, (V - R) = -0.03 \pm 0.03, (V
%- I) = 0.04 \pm 0.03$. %, (I - K) = -0.43 \pm 0.10.

The CMDs in Figure~\ref{fig-cmd} indicate the source lies very close
to the main sequence in this region of the LMC.  There appears to be a
slight excess of flux in the $I$ passband, as can be seen in the
$(V-R), (R-I)$ and $(V-I), V$ diagrams.  There is no apparent excess
in the $(V-R), V$ diagram, suggesting this is an infrared excess.
Given the direction of the reddening vector in the $V, R$, and $I$
passbands, this marginal excess cannot be a feature of reddening.
This could possibly be due to the lens itself, or in the context of
model X1, a signature of the dark companion to the primary.  We also
indicate in Figure~\ref{fig-cmd} the region 0.75 mag {\it below} the
lensed object with an open circle, which would be the location of a
single component of this object if it were a blend of equal brightness
stars, as model X2 suggests.

%It appears from the $(V-R), V$ CMD in Figure~\ref{fig-cmd} that the
%source is not significantly redder than the main sequence at this
%location in the LMC, implying the source is not on the far side of the
%LMC \cite{zhao-red}.  The lensed source appears to be on the main
%sequence, indicating it is unlikely the source is comprised of 2 stars
%of similar color and brightness, as fit X2 indicates.  Such an object
%should lie 0.75 magnitudes above the main sequence.  However, this
%constraint is weaker than we would like due to the steep slope of the
%main sequence in Figure~\ref{fig-cmd}, showing little color evolution
%with brightness.  We indicate in Figure~\ref{fig-cmd} the region 0.75
%mag {\it below} the lensed object with an open circle, which would be
%the location of a single component of this object if it were a blend
%of equal brightness stars.  This seems incompatible with the stellar
%distribution in the $V,(V-I)$ CMD, although this CMD also shows
%significant scatter.

\subsubsection{Properties of the Binary System}
\label{subsec-binproperties}

We next attempt to estimate the mass of an object with the colors
determined above.  Since this object appears on the upper main
sequence, it is likely to have higher metallicity than the majority of
the field stars in this location.  Here we consider
\citeN{girardi2000} isochrones for $z=0.008$.  For model class X1, the
best fits to the colors and apparent magnitude of the star come from
objects with $\log({\rm age[yr]}) = 8.7 - 8.8$, $M = 2.1 \pm 0.1
\msun$.  In the case of model X2, we require a single object $0.75$
mag dimmer than our observed object.  Using the same isochrones, we
find a wider range in acceptable star age, $\log({\rm age[yr]}) = 8.5
- 8.8$, and a range in mass of $M = 1.9 \pm 0.1 \msun$.

The original fits to model X1 yielded $\tilde{m}_1 \sim \tilde{f}_1
\sim 0$.  In the case of $\tilde{f}_1 = 0$, we are not able to fit
model parameters $\tilde{m}_1$ and $\tilde{\rho}$ individually, but
can only constrain their product $\tilde{m}_1~*~\tilde{\rho}$.
Therefore the fitting process was re--run, setting $\tilde{m}_1 \equiv
0.1$ ($\tilde{m}_1 = 0$ or $\tilde{m}_1 = 1$ imply non--existent
secondaries) and fitting for $\tilde{\rho}$.  Therefore fit X1 has 1
more degree of freedom than X2.  The product
$\tilde{m}_1~*~\tilde{\rho}$ was similar to that found in the original
model.  To explore the range of companions to the $2.1 \msun$ primary,
we consider a ``light'' dwarf secondary of $0.1 \msun$, henceforth fit
X1a, and a ``heavy'' white dwarf or neutron star secondary of $1.4
\msun$, fit X1b.  Making these assumptions fixes $\tilde{m}_1$ and
allows us to extract an associated $\tilde{\rho}$.  The resulting
parameters $\tilde{m}_1$ and $\tilde{\rho}$ for fits X1a and X1b are
also listed in Table~\ref{tab-xalparams}.

Knowing the total mass of the system and orbital period allows us to
solve for the physical Keplerian parameters of the binary source.  We
find semi-major axes of 0.11, 0.13, and 0.23 AU, and circular
velocities of 132, 154, and 120 $\kms$, for fits X1a, X1b, and X2,
respectively.

\subsubsection{Constraints on the Lensing Object}
\label{subsec-constraints}

Xallarap fit parameter $\tilde{\rho}$ relates the scale of the lens's
projected Einstein radius ($\Rehat$) to the binary's semi-major axis.
The derivation above allows us to express this value in AU, and the
results are presented in Table~\ref{tab-lens}.  This effectively leads
to a measurement of the lens proper motion

\begin{equation}
%%%\mu = {\vperp\over D_{l}} = {2 \Re^{'}\over \that D_{s}}.
\mu = {\vperp\over D_{l}} = {\Rehat\over t_{\rm E} D_{s}}.
\end{equation}

To find the velocity of the lens projected to the LMC ($\vhat$), we
assume a LMC distance modulus of 18.5, or a distance of 50 kpc.  We
further assume the source is 1 scale height behind the midplane of the
LMC disk (see below).  The values of $\mu$ and $\vhat$ for each of our
fits are presented in Table~\ref{tab-lens}.  The error bars on these
values do not incorporate uncertainty in the distance to the LMC.

The proper motion measurement allows a one-parameter family of
solutions relating the lens mass and distance

\begin{equation}
\label{eq-mlens}
%%%\mlens = {\mu^2 \that^2 c^2\over 16 G}
%%%          {D_{s} x\over 1 - x}  \ .
\mlens = {\mu^2 t_{\rm E}^2 c^2\over 4 G}
          {D_{s} x\over 1 - x}  \ ,
\end{equation}

where $x$ is the ratio of lens to source distances, $D_{l} / D_{s}$.

With some knowledge of the kinematics of sources and lenses, we can
estimate a most likely distance for the lensing object, given our
measured $\vhat$.  This directly leads to an estimate of the lens mass
from Equation~(\ref{eq-mlens}).  This is performed in a manner similar
to \citeN{macho-apj97e} and \citeN{macho-binary} for each of the
results from models X1a, X1b, and X2.  We use the following
representation of the Galactic-LMC system: the Galactic halo is
represented by Model S of \citeN{macho-apj97d}; we use the preferred
LMC disk model of \citeN{gyuk99}, with a scale-height of 300 pc, tilt
of 30~$\deg$ from face-on, central surface density of 190
$\msun$/pc$^2$, and we assume the source to be 1 scale height behind
the midplane; the Solar velocity in Galactic X, Y, Z is (9, 231, 16)
$\kms$; the LMC velocity is (53, -160, 162) $\kms$, with isotropic
random velocities of 22 $\kms$ (e.g., \citeNP{graff-carbon}) in each
direction.  For the LMC halo, we assume a similar profile to the
Galactic halo, with a central density of $0.0223 \msun {\rm pc}^{-3}$,
core radius of $2 \kpc$, and velocity dispersion of $50 \kms$.  We
truncate this halo at $11 \kpc$.  We note the details of the Galactic
and LMC models do not qualitatively alter our conclusions.

%%%With some knowledge of the kinematics of sources and lenses, we can
%%%estimate a most likely distance for the lensing object, given our
%%%measured $\vhat$.  This directly leads to an estimate of the lens mass
%%%from Eq.~\ref{eq-mlens}.  This is performed in a manner similar to
%%%\citeN{macho-apj97e} and \citeN{macho-binary} for each of the results
%%%from models X1a, X1b, and X2.  We use the following representation of
%%%the Galactic-LMC system: the Galactic halo is represented by Model S
%%%(Eqn.9) of \citeN{macho-apj97d}; the LMC is modeled as exponential
%%%disk with a scale-height of 250 pc, tilt of 20~$\deg$ from face-on,
%%%central face-on surface density of 363 $\msun$/pc$^2$, and the source
%%%is assumed 1 scale height behind the midplane; the Solar velocity in
%%%Galactic X, Y, Z is (9, 231, 16) $\kms$; the LMC velocity is (53,
%%%-160, 162) $\kms$, with isotropic random velocities of 22 $\kms$
%%%(e.g., \citeNP{graff-carbon}) in each direction.

\begin{figure}
\plotone{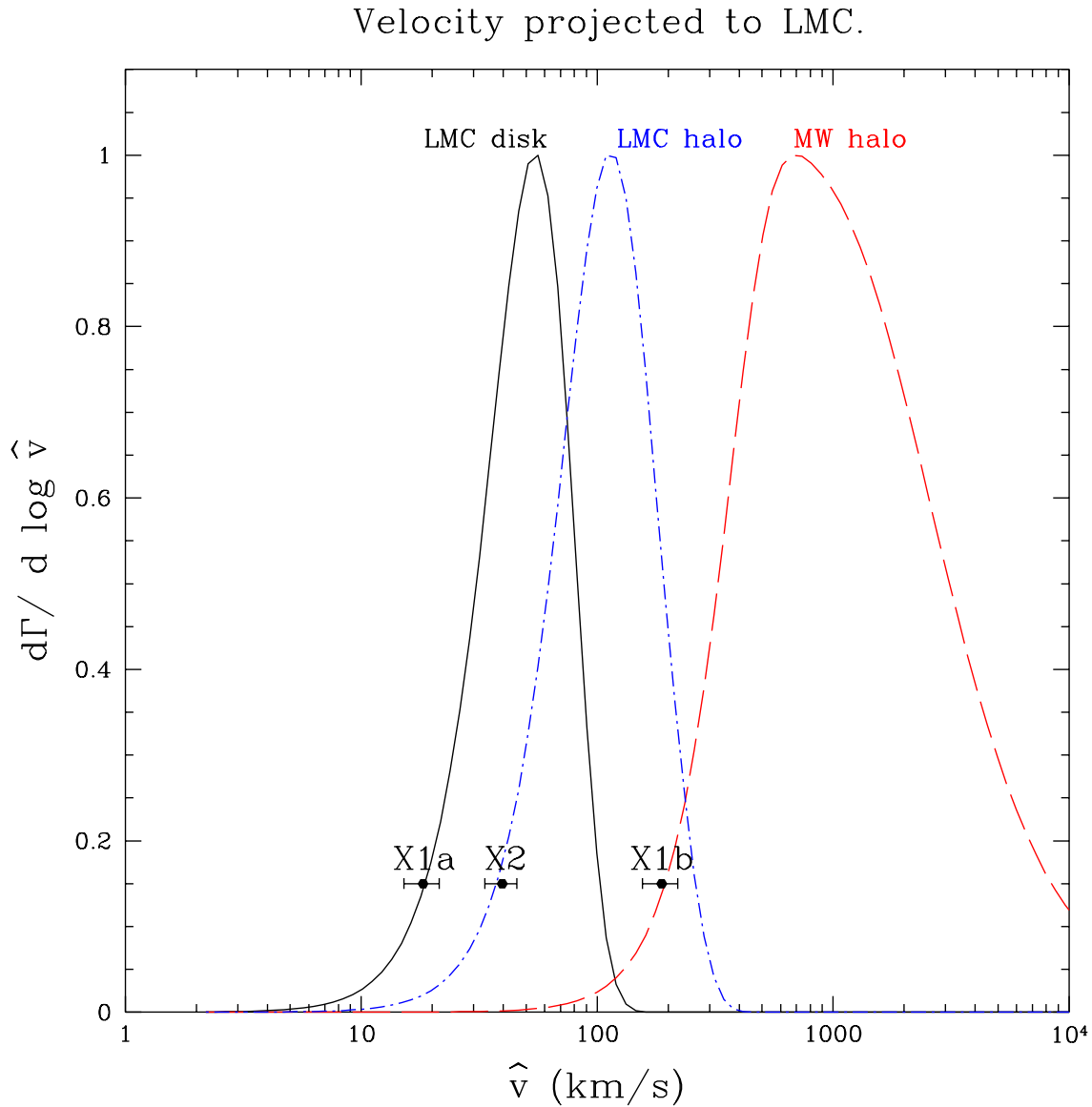}
\figcaption[f6.ps]{\label{fig-like}
  Probability distributions of $\vhat$ for each component of our
Galactic-LMC model, described in Sec.~\ref{subsec-constraints}.  The
measured $\vhat$ for models X1a, X1b, and X2 is also indicated, with
respective $1-\sigma$ errors.  Note in each case, a LMC lens is
preferred over a Galactic halo lens, although this preference is
weakest for model X1b.
}
\end{figure}

We find in each analysis that a lens residing in the LMC is preferred
to a Galactic halo lens, although only marginally so in fit X1b.  In
particular, for fit X1a the likelihood of measuring our value of
$\vhat_{X1a} = 18.3 \kms$ is dominated by the LMC disk model.  For
this fit, a LMC disk lens is 7 times more likely than a LMC halo lens,
and $> 950$ times more likely than a Galactic halo lens.  We also
point out that secondaries less massive than $0.1 \msun$ become
increasingly unlikely, as decreasing the mass of the secondary leads
to a lower lens $\vhat$.  In this model, $\vhat_{X1a}$ already is
drawn from the low velocity tail of the LMC disk probability
distribution.  As an example, a secondary of approximately Jupiter
mass would imply a lens velocity projected to the LMC of $\sim 0.2
\kms$, clearly in contradiction to the LMC disk likelihood profile in
Figure~\ref{fig-like}.  For fit X1b, we find $\vhat_{X1b} = 188 \kms$,
which is most likely to come from our model of the LMC halo, although
it is only 4 times as likely as a Galactic halo lens.  In this case, a
LMC disk lens is ruled out at high confidence.  We note that such a
LMC halo population has yet to be detected directly.  Given the broad
range in secondary mass, and hence the binary's semi-major axis,
explored between fits X1a and X1b, we conclude this model favors a
lens associated with the LMC.  With our model X2, we find $\vhat_{X2}
= 39.6 \pm 6.1 \kms$, whose likelihood is strongly dominated by the
LMC disk. Figure~\ref{fig-like} shows the probability distributions of
$\vhat$ for each component of our Galactic-LMC model.  Our measured
$\vhat$ for each fit is also displayed with $1-\sigma$ errors.  Only
in model X1b does a Galactic halo lens appear reasonable.

Our likelihood analysis also yields a probable distance to the lens,
and from Equation~(\ref{eq-mlens}), a mass.  These parameters are
listed in Table~\ref{tab-lens} for each model.  Model X1a allows the
lightest lens mass, $0.057{+0.060\atop -0.036} \msun$ , while models
X1b and X2 imply heavier lenses, $0.24{+0.38\atop -0.18}$ and
$0.28{+0.33\atop -0.20} \msun$, respectively.  We have referenced the
low mass isochrones of \citeN{girardi2000} to determine the expected
brightness of a main sequence M-dwarf lens.  We use the isochrones for
$z=0.004$, and an assumed lens age of 4 Gyr.  We find for objects of
$M = 0.3 \msun$, and also for the upper $1-\sigma$ confidence limit $M
= 0.6 \msun$, absolute V magnitudes of 9.9 and 7.6, respectively.  At
the distances implied by the likelihood analysis, even the most
luminous configuration provides negligible flux from the lens compared
to the apparent brightness of the source object.  A $0.6 \msun$ lens
at $45 \kpc$ leads to an apparent lens magnitude of $V = 25.9$, or
$0.2\%$ of the source brightness.  The apparent I--band brightness of
this lens is $I = 24.7$, or less than $0.1\%$ of the source brightness
in this band.  The secondary source is therefore a more likely origin
for the possible I--band excess seen in Figure~\ref{fig-cmd}.  The
small lens proper motion of $\lesssim 1$ milli-arcsecond year$^{-1}$
implies it will take the next generation of space telescopes to be
able to resolve and image the lensing object.

%X2  implies 0.3 at 49.6 kpc, 0.6 at 49.95 kpc.
%          dist mod 18.5             18.5
%           appar m 28.4             26.1
%
%X1b implies 0.3 at 38.9 kpc, 0.6 at 44.97 kpc.
%          dist mod 17.9             18.3
%           appar m 27.8             25.9

We note that the lens' location (and therefore its mass) can be more
accurately determined with direct spectral observations of the source.
In particular, the superposition of primary and secondary spectra can
provide a discriminant between models X1 and X2, and lead to a better
mass estimate for each object.  Similarly, the orbital period can be
precisely determined with radial velocity measurements.  These
observations will constrain the Keplerian parameters more directly
than our (necessarily) more complicated microlensing analysis has
allowed.

\subsection{Alternate Models}

In order to gauge the robustness of these conclusions, we have
considered several alternate models, including placing the sources at
larger distances behind the LMC, and increasing the amount of
reddening to the sources.

\citeN{zhao-extinct} predicts a strong excess in lensed source
reddening for models where the LMC is strongly self--lensing.  In
these models, the source stars must preferentially lie at the back
side of the LMC, and thus will have high interstellar reddenings.  On
the other hand, if the lenses are Galactic dark matter, the reddenings
of the source stars will be statistically the same as surrounding
stars along the same lines of sight.  \citeN{macho-lmcself00} use HST
colors of 8 microlensed source stars to rule out, at the 85\%
confidence level, a model where all sources are located $\sim$7 kpc
behind the LMC disk.  However, it is possible that roughly half of the
sources may be in the background.  Therefore, we consider a model
where the lensed sources are $7-9 \kpc$ behind the LMC.  This leads to
an increase in source mass of $\sim 0.1 \msun$, and results in
projected velocities of $\vhat_{X1a} = 18 \kms, \vhat_{X1b} = 184
\kms$, and $\vhat_{X2} = 40 \kms$.  The $\vhat$ likelihood profiles
for this source location differ significantly from those seen in
Figure~\ref{fig-like}.  Most notably, the LMC disk and LMC halo
profiles allow a similar range in $\vhat$, with the LMC halo generally
preferred over the LMC disk, and median $\vhat \sim 100 \kms$.  For
the lowest projected velocity (model X1a), the lens is consistent with
neither the LMC nor the Galactic likelihood profiles.  The LMC halo is
the most likely location of the lens in all models, followed by the
LMC disk in model X2, and by the Galactic halo in model X1b.

It is also possible the reddening to the sources is larger than our
adopted value of $\Ebv = 0.07$.  For example, the foreground reddening
maps of \citeN{schwering-91} indicate $\Ebv = 0.12$, while
\citeN{zartisky-red99} finds typical LMC source star reddenings of
$\Ebv \sim 0.10$.  If we assume a higher reddening to the source
object of $\Ebv = 0.12$, its intrinsic brightness and colors are $V =
19.06, (V - R) = -0.08, (V - I) = -0.05$.  For model X1, the primary
source would weigh $\sim 2.3 \msun$, and for model X2 both sources
would weigh $\sim 2.1 \msun$.  Here we find $\vhat_{X1a} = 17 \kms,
\vhat_{X1b} = 181 \kms$, and $\vhat_{X2} = 41 \kms$, whose relative
likelihoods can be evaluated using Figure~\ref{fig-like}.  This change
has more important implications for the source mass than for the
projected velocities, and the likelihood results are similar to those
from the standard analysis.

\section{Conclusions}

We have measured and characterized a periodic modulation in the
lightcurve of microlensing event MACHO 96-LMC-2.  We model this event
using a single object microlensing a rotating binary source, which
provides considerable improvement over a single source model.
Possible alternate explanations to this modulation include a single
source lensed by a rapidly rotating binary lens, or some variant of
stellar pulsation.  We do not consider such models.  MACHO 96-LMC-2 is
not the only time that binary source effects have been detected in a
microlensing event.  EROS-II event GSA2 \cite{eros-gsa} exhibits a
similar modulation around the standard microlensing fit.
\citeN{eros-gsa} also find a binary source model degeneracy similar to
that between our models X1 (dominant source) and X2 (equal brightness
sources).

We are able to constrain the projected Einstein radius of the lens
($\Rehat$) in the most significant of our fits to be between 0.54 and
5.5 AU, and its velocity projected to the LMC $\vhat$ to be between
18.3 and 188 $\kms$.  The weakness of this constraint (an order of
magnitude!) is due to our lack of knowledge of the mass of the
secondary component of the lensed binary system.  In this model we
have no means to constrain the secondary's orbit due to its negligible
contribution to the system's brightness.  We chose example secondaries
separated by an order of magnitude in mass ($0.1 \msun$ to $1.4
\msun$), leading to order of magnitude constraints on $\Rehat$ and
$\vhat$.  However, in both cases, a LMC lens is preferred to a
Galactic halo lens.  For the larger velocity lens (model X1b) the
object should come from an as yet undetected LMC halo population, and
a Galactic halo lens is not strongly ruled out.  The lack of direct
evidence for an LMC halo population indicates model X1b is best able
to constrain the location of the lens to be out of the LMC disk.  Our
second most significant model leads to $\Rehat = 1.24 \pm 0.18$ AU,
and $\vhat = 39.6 \pm 6.1 \kms$.  This model also prefers a lens in
the LMC, with the likelihood heavily weighted towards the LMC disk.
All 3 of these models suggest a sub-solar mass lens, consistent with
an M-dwarf star.  Our derived values of $\Rehat$, $\vhat$, $\mlens$,
and $D_{l}$ are in good agreement with the characteristic properties
of LMC lenses presented in Table~1 of \citeN{han97a}.

Alternate models for the source system, including one where the
sources lie $7-9 \kpc$ behind the LMC, and one where they are reddened
by an additional $\Delta \Ebv = 0.05$, were also considered.  By
placing the sources far behind the LMC, we are unable to discriminate
between the LMC disk and LMC halo population of lenses, and can only
state that the lens is most likely associated with the LMC system.
Additionally, the qualitative conclusions about the location of the
lens are found not to be overly sensitive to the amount of reddening
to the sources.

The identification of MACHO 96-LMC-2 with a LMC lens population is
consistent with the expected LMC self-lensing signal.  This single
event microlensing optical depth is $\tau_{A} = 1.1 \times 10^{-8},
\tau_{B} = 8.5 \times 10^{-9}$ for criteria ``A'' and ``B'' in
\citeN{macho-lmc5}, respectively.  This is to be contrasted with an
expected self-lensing optical depth of $\tau = 1.6 \times 10^{-8}$ for
the LMC disk and $\tau = 1.7 \times 10^{-8}$ for the MACHO component
of our model LMC halo \cite{macho-lmc5}.  It is also possible that
MACHO LMC-9 is due to LMC self-lensing \cite{macho-binary}, although
we caution this interpretation is based on a caustic crossing resolved
with only 2 observations, and other interpretations are possible.
LMC-9 is excluded from event set A in \citeN{macho-lmc5}, and has an
optical depth of $\tau_{B} = 9.3 \times 10^{-9}$.  The expected LMC
self-lensing signal due to these 2 events is thus likely to lie within
the range $1.1 \times 10^{-8} < \tau < 1.8 \times 10^{-8}$.  This is
approximately half of the expected LMC self-lensing rate.  Thus the
combined optical depths alone do not implicate the LMC as the host for
the majority of the microlenses, as originally suggested by
\citeN{sahu94a}.

\citeN{kerins-evans} take the ensemble of events LMC-9 and 98-SMC-1
and argue that the existence of even a few LMC self-lensing events
suggests an LMC halo interpretation of all LMC microlensing.  In this
context, the appearance of another apparent LMC self-lensing event
strengthens the case for significant LMC self-lensing.  However,
because of the importance of our result for the interpretation of LMC
microlensing, we emphasize the bias a study such as this has against
revealing a lens residing in our Galactic halo.  In particular, a
lensed LMC binary source is preferentially more likely to show
xallarap modulations if the lens is also in the LMC.  To identify
Galactic halo lenses, a spectroscopic study of {\bf all} Magellanic
Cloud microlensed sources should be undertaken to assess whether or
not the lensed object is, in fact, a binary system.  In the case of a
positive detection, we may set limits on xallarap modulation in the
microlensing lightcurve, and from this a lower limit on the proper
motion of the lensing object.  As can be seen from
Figure~\ref{fig-like}, Galactic halo lenses start to dominate the
likelihood near $\vhat = 230 \kms$.  A lower limit on $\vhat$ near
this value would be highly suggestive of a true halo lens, and thus
detection of at least one component of the Galactic dark matter.  We
suggest such a study of the high magnification event MACHO 99-LMC-2
\cite{99lmc2-iauc}, whose lightcurve is well covered by the
MACHO/GMAN, MOA, MPS, and OGLE microlensing teams.

\section*{Acknowledgments}

We are very grateful for the skilled support given our project by
S.~Chan, S.~Sabine, and the technical staff at the Mt. Stromlo
Observatory.  We especially thank J.D.~Reynolds for the network
software that has made this effort successful.  We would like to thank
the many staff and observers at CTIO and UTSO who have helped to make
the GMAN effort successful.

This work was performed under the auspices of the U.S. Department of
Energy by University of California Lawrence Livermore National
Laboratory under contract No. W-7405-Eng-48.  Work performed by the
Center for Particle Astrophysics personnel is supported in part by the
Office of Science and Technology Centers of NSF under cooperative
agreement AST-8809616.  Work performed at MSSSO is supported by the
Bilateral Science and Technology Program of the Australian Department
of Industry, Technology and Regional Development.  DM is also
supported by Fondecyt 1990440.  CWS thanks the Packard Foundation for
their generous support.  WJS is supported by a PPARC Advanced
Fellowship. CAN was supported in part by an NPSC Fellowship.  KG was
supported in part by the DOE under grant DEF03-90-ER 40546.  TV was
supported in part by an IGPP grant.  Support for this publication was
provided by NASA through proposal number GO-7306 from the Space
Telescope Science Institute, which is operated by the Association of
Universities for Research in Astronomy, under NASA contract
NAS5-26555.

\clearpage

%\bibliography{apjmnemonic,general_refs,macho_refs,binary_refs,myrefs}
%\bibliography{apjmnemonic,newbib}
%\bibliographystyle{apj}

\clearpage

\clearpage

\begin{deluxetable}{cll}
\tablecaption{Standard Microlensing Fit Parameters \label{tab-stdparams} }
\tablewidth{0pt}
\scriptsize
\tablehead{
  \colhead {Fit Parameter} &
  \colhead {} 
}
\startdata
$\chi^2$ / d.o.f.       %2073 points, 13 pars
%%   & 1877.36
   & 1872.31 / 2060 \nl
$\t0$ \tablenotemark{a} 
%%  & 1767.50 (7)
   & 1767.51 (7) \nl
$\that$ 
%%   & 98.85   (58)
   & 98.7    (40) \nl
$\umin$
%%  & 0.300   (1) 
   & 0.301   (17)\nl
$f_{MACHO_R}$
%%   & 1 
   & 1.00 (13)  \nl
$f_{MACHO_B}$
%%   & 1 
   & 1.00 (13) \nl
$f_{CTIO_R}$ 
%%   & 1 
   & 0.99 (12) \nl
$f_{CTIO_B}$
%%   & 1 
   & 1.02 (13) \nl
$f_{UTSO_R}$
%%   & 1 
   & 0.76 (33) \nl
\enddata
\tablenotetext{a} { (JD $-$ 2448623.50). }
\tablenotetext{} { Standard point source, point lens microlensing fit
parameters for event 96-LMC-2.  For each passband, $f$ represents the
fraction of the object's baseline brightness which was lensed.
Reported uncertainties in the final significant digit(s) are the
maximum extent of the surface in parameter space which has a $\chi ^2$
greater than the best--fit value by 1.}
\end{deluxetable}

\begin{deluxetable}{cllll}
\tablecaption{Xallarap Microlensing Fit Parameters \label{tab-xalparams} }
\tablewidth{0pt}
\scriptsize
\tablehead{
  \colhead{} &
  \colhead{X1}  &
  \colhead{X1a}  &
  \colhead{X1b}  &
  \colhead{X2}  
}
\startdata
$\chi^2$ / d.o.f.        %2073 points, 20 pars
   & 1799.9 / 2054
   & \nodata
   & \nodata
   & 1800.9 / 2053   \nl
\hline
$\tilde{t}_b$ \tablenotemark{a} 
   & 1767.600 (79)
   & \nodata
   & \nodata
   & 1767.80 (14)          \nl
$2*t_{\rm E}$
   & 101.8 (42)
   & \nodata
   & \nodata
   & 108.8 (53)          \nl
$\tilde{b}$
   & 0.287 (16)
   & \nodata
   & \nodata
   & 0.246 (19)          \nl
$\tilde{\alpha}$
   & -2.15 (37)
   & \nodata
   & \nodata
   & 0.236 (77)          \nl
$\tilde{f}_1$
   & 0.000 (29)
   & \nodata
   & \nodata
   & 0.52 (19)          \nl
$\tilde{m}_1$
   & 0.1 \tablenotemark{b} 
   & 0.045
   & 0.40
   & 0.53 (19)          \nl
$\tilde{\rho}$
   & 0.095 (15)  \tablenotemark{b} 
   & 0.208 (33)
   & 0.0237 (39)
   & 0.188 (26)         \nl
$\tilde{\beta}$
   & 1.00 (30)
   & \nodata
   & \nodata
   & 1.43 (14)          \nl
%$\tilde{\gamma}$
%   & -0.0560 (8)
%   & \nodata
%   & \nodata
%   & 0.59 (66)          \nl
$\tilde{T}$
   & 9.22 (21)
   & \nodata
   & \nodata
   & 21.22 (53)          \nl
$\tilde{\xi}_0$
   & -0.23 (37)
   & \nodata
   & \nodata
   & 0.40 (12)          \nl
\hline 
$f_{MACHO_R}$
   & 0.94 (9)
   & \nodata
   & \nodata
   & 0.80 (8)          \nl
$f_{MACHO_B}$
   & 0.95 (9)
   & \nodata
   & \nodata
   & 0.80 (8)          \nl
$f_{CTIO_R}$ 
   & 0.93 (9)
   & \nodata
   & \nodata
   & 0.79 (8)          \nl
$f_{CTIO_B}$
   & 0.96 (9)
   & \nodata
   & \nodata
   & 0.81 (8)          \nl
$f_{UTSO_R}$
   & 0.58 (18)
   & \nodata
   & \nodata
   & 0.86 (46)          \nl
\enddata
\tablenotetext{a} { (JD $-$ 2448623.50). }
\tablenotetext{b} { For fit X1, we are only able to constrain the
product of $\tilde{m}_1$ and $\tilde{\rho}$.  We assume $0.1 \msun$
and $1.4 \msun$ dark companions to the $2.1 \msun$ primary to
determine these parameters for fits X1a and X1b, respectively.  The
sources for fit X2 are estimated to be $1.9 \msun$.  See
Sec.~\ref{subsec-binproperties} for further details. }
\tablenotetext{} { Xallarap microlensing fit parameters for event
96-LMC-2.  The parameters are as defined in Sec.~\ref{subsec-xal}.
For each passband, $f$ represents the fraction of the objects baseline
brightness which was lensed.  Reported uncertainties in the final
significant digit(s) are the maximum extent of the surface in
parameter space which has a $\chi ^2$ greater than the best--fit value
by 1.}
\end{deluxetable}

\begin{deluxetable}{llll}
\tablecaption{Properties of the Lensing Object \label{tab-lens} }
\tablewidth{0pt}
\scriptsize
\tablehead{
  \colhead{} &
  \colhead{X1a} &
  \colhead{X1b} &
  \colhead{X2} 
}
\startdata
$\Rehat$~~(AU)
   & 0.535 (88)
   & 5.50 (91)
   & 1.24 (18) \nl
$\mu~~(\kmskpc)$
   & 0.364 (62)
   & 3.73 (64)
   & 0.79 (12) \nl
$\vhat~~(\kms)$
   & 18.3 (31)
      %%% prob lmcdisk   lmchalo   mwhalo
      %%%      0.1432530 0.0202372 0.0001505
   & 188 (32)
      %%% prob lmcdisk   lmchalo   mwhalo
      %%%      0.0000032 0.5277663 0.1378715
   & 39.6 (61) \nl
      %%% prob lmcdisk   lmchalo   mwhalo
      %%%      0.7795788 0.1740024 0.0015164
$\mlens~~(\msun)$
   & $5.7{+6.0\atop -3.6} \times 10^{-2}$   %0.057259{+0.060438\atop -0.035572}
   & $2.4{+3.8\atop -1.8} \times 10^{-1}$   %0.242266{+0.384649\atop -0.180763}
   & $2.8{+3.3\atop -2.0} \times 10^{-1}$ \nl %0.277751{+0.328958\atop -0.198408}
$D_{l}~~(\kpc)$ 
   & $49.6{+0.3\atop -1.0}$ 
   & $38.5{+6.5\atop -15.7}$
   & $49.6{+0.4\atop -1.6}$ \nl
\enddata
\tablenotetext{} { Characteristics of the lensing object, determined
from our xallarap fit parameters and estimates of the individual
masses of the lensed binary system.  $\Rehat$ represents the lens'
Einstein ring radius projected to the source system, and $\vhat$
represents the lens velocity projected to the source system.  Lens properties 
$\mlens$ and $D_{l}$ are estimated using a maximum likelihood technique
described in Sec.~\ref{subsec-constraints}, and a LMC distance of 50 kpc.}
\end{deluxetable}

\end{document}